\documentclass[doublecol,amsmath,amssymb]{epl2-kut} 
\usepackage{graphicx}
\usepackage{bm}

\def\timeevol#1{\setbox1=\hbox{$\longmapsto$}\setbox2=\hbox{$
  \scriptstyle #1$}\copy1\kern-.5\wd1\kern-.5\wd2
  \raise-1.2\ht2\copy2\kern-.5\wd2\kern\wd1}
\newcommand{\Tp}{T_{\rm p}}

\newcommand{\Tsymbol}{\mathop{\mathcal{T}}_{\leftarrow}}
\newcommand{\ATsymbol}{\mathop{\mathcal{T}}_{\rightarrow}}

\title{New Anatomy of Quantum Holonomy}

\author{
Taksu Cheon\inst{1} and 
Atushi Tanaka\inst{2}
}

\institute{                    
\inst{1}
Laboratory of Physics,
Kochi University of Technology,
Tosa Yamada, Kochi 782-8502, Japan\\
\inst{2}
Department of Physics,
Tokyo Metropolitan University, Hachioji, 
Tokyo 192-0397, Japan
}

\abstract{
We present a unified formulation of quantum holonomy
which is capable of describing all of its known varieties.
The full role of non-Abelian gauge connection is elucidated.
As examples, solvable models of quantum kicked spin are analyzed.
They are shown to exhibit spiral holonomy, as well as Berry phase 
and Wilczek-Zee holonomy. 
}

\date{November 17, 2008}

\begin{document}

\maketitle

The interest in Berry's geometric phase is now two-fold.
Theoretical interest to this quintessentially quantum phenomenon is 
well documented \cite{WS89}.  
The central importance is in the appearance of  the gauge structure
in parametric space  \cite{MT79, BE84, WZ84}.
Immediate utility of geometric phase
is recently highlighted by the suggestion of holonomic and
adiabatic quantum computings \cite{ZR99, JV00, FG01, TN05, ZH06}.  
%
Consider the change of a stationary quantum state
by an adiabatic variation of a system parameter.  When the variation of
the parameter is cyclic, namely it comes back to the original value, the state
does not necessarily returns exactly to the original stationary state.
This phenomenon, known as quantum holonomy, comes in
three flavors, Berry phase, Wilczek-Zee holonomy, and spiral eigenvalue
holonomy.
%
In case of Berry phase, 
a stationary state comes back to itself with extra phase added
after an adiabatic cyclic parameter variation, 
and in case of
Wilczek-Zee holonomy, a state belonging to a set of degenerate 
eigenstates turns into a mixing of those states sharing the same energy.
In case of spiral holonomy, on the other hand, an eigenstate evolves into
another eigenstate with {\it different energy} 
after an adiabatic cyclic parameter variation \cite{CH98, TF01, TM07, MT07}.
%
%
Compared to the well-known first two,
which has been thoroughly studied,  
the newly-found spiral holonomy
has been something of an odd man out,
since it is yet to find its proper place in the
general formulation of quantum holonomies in terms of gauge connections,
and this might be the reason for the lack of due recognition and
appreciation of the existence of the spiral holonomy.
%
This is deplorable
for practical reason  in understanding the mechanism of
spiral holonomy,
as well as it is theoretically unsatisfactory,
since spiral holonomy
has obvious advantages in both control robustness
and variety of achievable states with adiabatic quantum-state control, 
which enhance its usefulness in quantum computing \cite{MT07}. 

In this paper, we intend to solve the problem of making sense of 
spiral holonomy, and prove that it indeed belongs to the
category of quantum holonomy.
We show that the non-Abelian Mead-Berry gauge connection is
again the key in
developing a unified formalism which treats all the quantum holonomy 
in a single fold.  
The resulting formula also sheds new light on the gauge invariance
of quantum holonomy and its implications.
%

%
%
We start by
considering a quantum system described by a
Hamiltonian which depends on sets of parameters,
which are collectively referred to as $\alpha$.
Let us assume that the parameter is temporally varied $\alpha=\alpha(t)=\alpha_t$.
Quantum evolution is described by the Schr{\" o}dinger equation
\begin{eqnarray}
\label{e101}
\left( H(\alpha_t) - i\frac{d} {d t} \right) \left| \Psi(t) \right> = 0 .
\end{eqnarray}
We look for the solution of (\ref{e101}) in terms of a set of arbitrary
``moving'' basis states 
$\{ \left| v_n(\alpha) \right> \}$ with time-dependent parameter $\alpha=\alpha_t$.
%
%
The temporal evolution of $\left| \Psi(t) \right>$ is described by 
%
\begin{eqnarray}
\label{e102}
\left| \Psi(t) \right> = U(t) \left| \Psi(0) \right> ,
\end{eqnarray}
where the evolution operator $U(t)$ is given by
a time-ordered exponential \cite{FU05} 
%
\begin{eqnarray}
\label{e103}
U(t) = 
\sum_{m,n}  \left| v_m(\alpha_t) \right>
\left[ \Tsymbol e^{-i\int^t_0 dt F(\alpha_t)} \right] _{m,n} 
\left< v_n(\alpha_0) \right| ,
\end{eqnarray}
where
$\Tsymbol$  signifies the temporal ordering of the operator integral 
from right to left, and 
the elements of the matrix $F(\alpha)$  are given by
%
\begin{eqnarray}
\label{e104}
\quad
F_{m,n}(\alpha) 
= 
\left< v_m(\alpha) \right| (H(\alpha) -{\dot \alpha} i \partial_\alpha) 
\left| v_n(\alpha) \right> .
\end{eqnarray}
%
In deriving (\ref{e103}), we have used a decomposition of the state
$\left| \Psi(t) \right>=\sum_m  \left| v_m(\alpha_t) \right> b_m(t)$,
and the evolution equation  $i {\dot b_m}(t) = \sum_n F_{m,n}(\alpha_t) b_n(t)$
which is derived from (\ref{e101}).
We further rewrite (\ref{e103}) into the form
\begin{eqnarray}
\label{e105}
U(t) =\!
\sum_{l,m,n}  \left| v_l(\alpha_0) \right>
W_{l,m}(\alpha_0,\alpha_t)D_{m,n}(t,0) 
\left< v_n(\alpha_0) \right| ,
\end{eqnarray}
%
with matrices $W$ and $D$ given by path-ordered and time-orderd
exponentials, respectively,
\begin{eqnarray}
\label{e106}
&&\!\!\!\!\!\!\!\!
W(\alpha_0,\alpha_t)= 
\ATsymbol e^{-i\int_{\alpha_0}^{\alpha_t} d\alpha A(\alpha)},
\nonumber \\ 
&&\!\!\!\!\!\!\!\!
D(t,0)= \Tsymbol e^{-i\int_0^t dt F(\alpha_t)},
\end{eqnarray}
where $\ATsymbol$  signifies the temporal ordering from left to right,
and
$A(\alpha)$ is the full {\it non-Abelian} Mead-Berry gauge connection whose
matrix elements are defined by
%
\begin{eqnarray}
\label{e107}
A_{m,n}(\alpha) =
\left< v_m(\alpha) \right| i {\partial_{\alpha}} \left. v_n(\alpha) \right> .
\end{eqnarray}
Note, in the derivation of (\ref{e105}), the  identity
$ \left| v_m(\alpha') \right> 
= \sum_l {\left| v_l(\alpha) \right> W_{l,m}(\alpha,\alpha')}$,
signifying the fact that $W$ is a displacement operator in parameter space.
Seemingly minor amendment of the factor $W$, in (\ref{e105}), 
to the Fujikawa's formula (\ref{e103}) is to be crucial  in the following arguments. 
Let us now assume  $\left| v_n(\alpha) \right>$ to be the ``instantaneous'' 
eigenstates of $H(\alpha)$ with fixed value of $\alpha$, namely
\begin{eqnarray}
\label{e108}
H(\alpha) \left| v_n(\alpha) \right> = \varepsilon_n(\alpha) \left| v_n(\alpha) \right> ,
\end{eqnarray}
and further assume the standard {\it adiabatic} approximation that amounts to suppose
the variation of $\alpha$ to be so slow that  the only significant contribution 
to (\ref{e106}) comes from the diagonal term in $F(\alpha_t)$;
\begin{eqnarray}
\label{e109}
F_{m,n}(\alpha_t)
\approx 
\delta_{m,n} \left\{ \varepsilon_n(\alpha_t) - {\dot \alpha}A_{n,n}(\alpha) \right\} .
\end{eqnarray}
We now consider a cyclic motion along $C: \alpha_0 \to \alpha_\tau$
with the identification of $\alpha_\tau$ with $\alpha_0$. 
We have
\begin{eqnarray}
\label{e110}
U(\tau) =
\sum_{m,n}  \left| v_m(\alpha_0) \right>
           M_{m,n}(C) e^{-i \phi_{d}}
           \left< v_n(\alpha_0)\right|  ,
\end{eqnarray}
where $\phi_{d}=\int_0^\tau dt \varepsilon_n(\alpha_t)$ is the dynamical phase.
The   {\it holonomy matrix} $M(C)$  encodes 
the geometrical change,
independent from the temporal rate of evolution, 
induced by the adiabatic cycle $C$,
and is given by
\begin{eqnarray}
\label{e211}
M(C) = W(C) B(C),
\end{eqnarray} 
in which each factor is written as
\begin{eqnarray}
\label{e212}
&&\!\!\!\!\!\!\!\!
W(C) = \ATsymbol e^{-i \oint_{C} d\alpha A(\alpha)} ,
\nonumber \\
&&\!\!\!\!\!\!\!\!
B(C) =  \Tsymbol e^{ i \oint_{C} d\alpha A^D(\alpha)}  ,
\end{eqnarray} 
where diagonal matrix $A^D$, defined as
%
\begin{eqnarray}
\label{e213}
A^D_{m,n}(\alpha) = A_{n,n}(\alpha) \delta_{m,n} ,
\end{eqnarray} 
leaves  $B(C)$ also diagonal. 
The equations (\ref{e211})-(\ref{e212}) constitute the central result of this paper.
Several comments are now in order.

%
The expression (\ref{e110}) is notable for its manifest gauge invariance.
This is easily seen by changing the phases of individual states, 
\begin{eqnarray}
\label{e214}
\left| v_n(\alpha) \right> \to  
e^{i g_n(\alpha)} \left| v_n(\alpha) \right>.
\end{eqnarray} 
We then have 
%
$W_{m,n}(C)  \to  
e^{-i g_m(\alpha_0) + i g_n(\alpha_\tau)} W_{m,n}(C)$,
and
$
B_{n,n}(C)  \to  
e^{-i g_n(\alpha_\tau) + i g_n(\alpha_0)} B_{n,n}(C)$,
%
giving
\begin{eqnarray}
\label{e215}
M_{m,n}(C)  \to  
e^{-i g_m(\alpha_0) + i g_n(\alpha_0)} M_{m,n}(C) ,
\end{eqnarray} 
which obviously leaves $U(\tau)$ unchanged.
Note that the original $U(t)$ given by (\ref{e105}) is 
invariant with full gauge transformation
\begin{eqnarray}
\label{e916}
\left| v_n(\alpha) \right> \to 
G \left| v_n(\alpha) \right>
=\sum_m \left| v_m(\alpha) \right> G_{m,n}(\alpha),
\end{eqnarray} 
which simply is an alternative 
statement of the basis $\left| v_n(\alpha)\right>$ 
being arbitrary.

%
%
The adiabatic time evolution ensures that 
the system that starts with an eigenvector $\left| v_n(\alpha_0) \right>$
stays continuously in an instantaneous eigenstate 
$\left| v_n(\alpha_{t}) \right>$, modulo its phase.
After the completion of the adiabatic cycle $C$, $\alpha$ returns 
to the initial point $\alpha_\tau=\alpha_0$.
However, 
there is no {\it a priori} reason to assume that the state of the system 
comes back to its initial one,
even apart from its phase.
In fact, counterexamples to the conventional assumption
have been found in the form of
the spiral anholonomies in \cite{CH98} and \cite{TM07}.
In general, the holonomy matrix $M(C)$ 
is a {\it permutation matrix},
which describes how each $\left| v_n(\alpha_0) \right>$ is transported
into its orthogonal state (or itself),
accompanied by a possible geometric phase factor.

%
In the absence of spiral holonomy, $W_{m,n} = 0$ for $ m\ne n$, 
the gauge can been chosen
to make the basis state $\left| v_n(\alpha)\right>$ single-valued, 
thus giving $W_{m,n} = \delta_{m,n}$,
and the usual single factor expression of
Berry phase
$M_{m,n} = \delta_{m,n} B_{n,n}$ is obtained.
%
Note, however, that the factor $W(C)$ is necessary for consistency
even for this case,  in order to ensure the
correct Berry phase with arbitrary choice of gauge $g(\alpha)$,
which leaves the basis states multi-valued,  
$\left| v_n(\alpha_\tau) \right> \ne \left| v_n(\alpha_0) \right>$.

%
In fact, a special choice of the gauge can turn $B(C)$ into
a unit matrix, 
$B_{m,n}(C) = \delta_{m,n}$.  This is obtained with the condition
%
$A^D(\alpha) = 0$,
%
which is nothing but the {\it parallel transport} condition \cite{ST76}.
The holonomy matrix, then, takes a single factor form
\begin{eqnarray}
\label{e902}
M(C) = \ATsymbol e^{-i \oint_C d\alpha A(\alpha)}.
\end{eqnarray}
%
With this privileged choice of the gauge, all adiabatic quantum 
holonomies are described as  a holonomy of an ordered basis.
This is a generalization of Fujikawa's picture, in which 
the Berry phase is attributed to the holonomy of  basis vectors,
instead of probability amplitudes \cite{FU05}.

%
The validity of the split of the unitary matrix $D$ to geometrical part $B$ and 
the dynamical phase is established whenever non-diagonal elements
of $F$ are negligible, and this can happen independent from the adiabaticity 
of parametric variations.
One such case occurs 
when $\left| v_n(\alpha_t) \right>$ is the solution to the full problem (\ref{e101}).
Then the geometric matrix $M$ can be extracted without the assumption 
of adiabaticity, and this is exactly 
the situation considered by Aharonov and Anandan \cite{AA87}. 
%
%
%
%

In the presence of degeneracies, we need to modify the preceding
argument by writing the instantaneous eigenbasis as
\begin{eqnarray}
\label{e416}
H(\alpha) \left| v_{nk}(\alpha) \right> 
= \varepsilon_n(\alpha) \left| v_{nk}(\alpha) \right> ,
\end{eqnarray}
where $k$ indexes the different states sharing a degeneracy.
The quantum holonomy after cyclic variation of parameters along 
the path $C$ is described formally by  the same $M(C)$ as
in (\ref{e211})-(\ref{e212}) with 
%
%
$A^D$ now defined by block-diagonal
\begin{eqnarray}
\label{e417}
A^D_{m j, n k}(\alpha) = A_{n j, n k}(\alpha) \delta_{m,n},
\end{eqnarray} 
which results in block-diagonal $B(C)$ that mixes degenerate states after
cyclic parametric variation.
%
%
%
%
%
%
%
Because of the presence of $W$, 
combination of all three types of quantum holonomy, Berry, Wilczek-Zee
and spiral, can arise in general.
%

The preceding formalism can be straightforwardly extended to 
periodically driven systems.
Consider a system whose Hamiltonian has periodic time dependence 
$H(t+\Tp, \alpha) = H(t, \alpha)$.
The stationary state $\left| v_n(\alpha) \right>$ of this system is
an eigenvector of 
the Floquet operator, i.e., the time evolution operator for a unit period $\Tp$,
\begin{eqnarray}
\label{e418}
U_{\Pi}(\alpha)=\Tsymbol e^{-i\int_0^{\Tp} dt H(t, \alpha)}
.
\end{eqnarray}
We have an eigenvalue equation
%
\begin{eqnarray}
\label{e419}
U_{\Pi}(\alpha) \left| v_n(\alpha) \right> 
= e^{-i \varepsilon_n(\alpha)\Tp} \left| v_n(\alpha) \right>
,
\end{eqnarray}
where a quasienergy $\varepsilon_n(\alpha)$ is defined 
only up to modulo $2\pi/\Tp$.
When the time scale of the variation of $\alpha=\alpha_t$
is far larger than the unit period, $\tau \gg \Tp$, 
we can speak about the adiabatic change of $\alpha$.
Just as before,  $\left| v_n(\alpha) \right>$ determines the gauge
connection $A_{m,n}(\alpha)$,
that is to be used to compute the holonomy matrix $M(C)$.

%
%
We now illustrate our arguments with an example of kicked spin \cite{TM07, MT07},
which is a variant of systems with rank-one periodic kick \cite{CO90, MS90}.
Consider a spin one-half subjected to constant and time-periodic
magnetic force;
\begin{eqnarray}
\label{520}
H = T \sigma_3 + 
\lambda V \sum_{n=-\infty}^{\infty} \delta(t-n-\frac{1}{2}) .
\end{eqnarray}
where $V$ is given by a linear combination of spin operators $\sigma_1$, $\sigma_2$ and 
$\sigma_3$.
We require the system to be $2\pi$-periodic with respect to the strength parameter $\lambda$,
which gives the condition $e^{i 2\pi V} = 1$.
%
%
With straightforward computations, 
$V$ is shown to have the form
\begin{eqnarray}
\label{e521}
V =  \frac{p}{2}+\frac{2-p}{2}\sum_{i=1}^3 b_i \sigma_i  ,
\end{eqnarray}
with integer $p$ and $ \sum_i b_i^2=1$.
Let us adopt the spherical parametrizaton of three dimensional 
unit vector $\{ b_i \}$
as $b_1=\sin\gamma\cos\xi$,  $b_2=\sin\gamma\sin\xi$ and
$b_3=\cos\gamma$.
%
%
The three parameters of the system $\lambda$, $\gamma$, $\xi$ form 
a three dimensional manifold
%
$
\{ \lambda, \gamma, \xi \} =  S^1 \times S^2
$
.
%
The temporal evolution during one period $\Tp = 1$ is described by a unitary operator
%
$U_{\Pi} = e^{-\frac{i T}{2}\sigma_3} e^{-i\lambda V} e^{-\frac{i T}{2}\sigma_3}$.
%
%
The problem is fully solvable with elementary functions, 
%
\begin{eqnarray}
\label{e524}
\varepsilon_n = \frac{p}{2}\lambda + (-1)^n E_{ \frac{2-p}{2}\lambda, \gamma, T} ,
\end{eqnarray}
%
%
\begin{eqnarray}
\label{e525}
&&
\left| v_0 \right> = 
\pmatrix{
e^{-\frac{i \xi}{2}} \cos Q_{ \frac{2-p}{2}\lambda, \gamma, T} \cr
e^{\frac{i \xi}{2}} \sin Q_{ \frac{2-p}{2}\lambda, \gamma, T}
 } ,
\nonumber \\
&&
\left| v_1 \right> = 
\pmatrix{
-e^{-\frac{i \xi}{2}} \sin Q_{ \frac{2-p}{2}\lambda, \gamma, T} \cr
e^{\frac{i \xi}{2}} \cos Q_{ \frac{2-p}{2}\lambda, \gamma, T}
 } ,
\end{eqnarray}
%
Here, $E$ and $Q$ are given by
\begin{eqnarray}
\label{e526}
E_{\lambda, \gamma, T} 
= \cos^{-1} {\left( \cos\lambda \cos T - \sin\lambda \sin T \cos\gamma \right)} ,
%
%
\nonumber \\
Q_{\lambda,\gamma,T}
= \frac{1}{2} \tan^{-1} {\left( \!  
   \frac{ \sin\lambda \sin\gamma } 
           {\cos\lambda \sin T + \sin\lambda \cos T \cos\gamma}  \! \right)}.
\end{eqnarray}
Note the relations
$Q_{\lambda+\pi,\gamma,T} = Q_{\lambda,\gamma,T}+\frac{\pi}{2}$
and also
$Q_{\frac{\pi}{2},\gamma+\pi,T} = Q_{\frac{\pi}{2},\gamma,T}+\frac{\pi}{2}$, and
$Q_{\lambda,\gamma+\pi,0} = Q_{\lambda,\gamma,0}+\frac{\pi}{2}$.
%
%
%
%
Also note anti-periodicity of $E$ function; 
$E_{\lambda+\pi,\gamma,T} = \pi-E_{\lambda,\gamma,T}$.
%
%
%
%
%
The gauge connections 
$A^{\alpha} 
= [ \left< v_n \right| i \partial_\alpha \left| v_m \right> ]$
are given by
\begin{eqnarray}
\label{e527}
&&
A^{\xi}  
=\frac{1}{2} (\Sigma_3 \cos\gamma - \Sigma_1\sin\gamma) ,
\nonumber \\
%
&&
A^{\gamma} 
=\partial_\gamma Q_{\frac{2-p}{2}\lambda,\gamma,T} \Sigma_2 ,
%
\nonumber \\
%
&&
{ A^{\lambda}} 
=\partial_\lambda Q_{\frac{2-p}{2}\lambda,\gamma,T} \Sigma_2 ,
\end{eqnarray}
where $\Sigma_i$ are Pauli spin matrices acting on the space of vectors 
$\left| v_0 \right> = \left[ \begin{array}{cc} 1 \cr 0\end{array} \right]$
and
$\left| v_1 \right> = \left[ \begin{array}{cc} 0 \cr 1\end{array} \right]$.
%
%
%
We have
\begin{eqnarray}
\label{e528}
&&\!\!\!\!\!\!\!\!\!\!
W(C^\xi) = -1, \quad
W(C^\gamma) =-1,
\nonumber \\
&&\!\!\!\!\!\!\!\!\!\!
W(C^\lambda) 
= \cos Q_{{(2-p)\pi},\gamma,T}
-i \Sigma_2 \sin Q_{{(2-p)\pi},\gamma,T}  ,
\end{eqnarray}
and
\begin{eqnarray}
\label{e529}
B(C^\xi)
= e^{ i {\pi}\cos\gamma \Sigma_3 },
\quad
B(C^\gamma)
= 1,
\quad
B(C^\lambda)= 1.
\end{eqnarray}
After the cyclic variation of $\alpha$ 
along the path $C_\alpha$ : $\alpha \rightarrow \alpha+2\pi$,  
the state $\left| v_n \right>$ is transformed to $M(C_\alpha)\left| v_n \right>$ with
\begin{eqnarray}
\label{e530}
M({C_\xi}) 
=  e^{ -i\pi(1- \Sigma_3 \cos\gamma) } ,
\quad
M({C_\gamma}) =-1,
\end{eqnarray}
for $T=0$,
which signify the existence of ``monopole'' at the center of parameter space $S^2$,
and are the classic example of Berry's phase. 
It is worthwhile to point out that, with our choice of gauge (\ref{e525}),
the Berry phase $M(C_\gamma)=-1$ is give not from 
the expected place $B(C_\gamma)$, but from $W(C_\gamma)$,
which is a good illustration of the convenience of gauge invariant expression 
(\ref{e211}).
For the cyclic path of $\lambda$, we have
\begin{eqnarray}
\label{e531}
M({C_\lambda}) 
=   \cos \frac{(2-p)\pi}{2}    - i  \Sigma_2 \sin\frac{(2-p)\pi}{2}    ,
\end{eqnarray}
which clearly shows 
the exchange of eigenstates after the cyclic variation of $\lambda$
for odd $p$, that leaves only the second term.
Corresponding eigenvalue anholonomy is visible in the expression,
$\varepsilon_n(\lambda+2\pi) = \varepsilon_{n+p}(\lambda) $
%
%
and we have the spiral holonomy with energy shifts $\Delta n(C_\alpha)$
given by
%
$\Delta n(C_\xi) = 0$, 
$\Delta n(C_\gamma) = 0$ and 
$\Delta n(C_\lambda) = p$,
%
from which, the meaning of $p$ as {\it winding number} is evident. 
%

As an example of degenerate holonomy, we now turn to
a time-periodic extension of magnetic spin 
with higher angular momentum \cite{AS89}.
Consider a spin three-half subjected to constant and time-periodic
quadrupole force;
\begin{eqnarray}
\label{e633}
H = T \tau_5 + 
\lambda V \sum_{n=-\infty}^{\infty} \delta(t-n-\frac{1}{2}) ,
\end{eqnarray}
where $V$ is given by a linear combination of spin operators $\tau_1$,
$\tau_2$, $\tau_3$, $\tau_4$ and $\tau_5$, 
where
\begin{eqnarray}
\label{e634}
&&
\tau_1 = 
\pmatrix{
 {\bf 0} &  i\sigma_2  \cr  -i \sigma_2 \! &{\bf 0} 
 } ,
\
\tau_2 = 
\pmatrix{
 {\bf 0} &  -i\sigma_1  \cr i \sigma_1 &{\bf 0} 
 } ,
\nonumber \\
&&
\tau_3 = 
\pmatrix{
{\bf 0} & I \cr I &{\bf 0}
 } ,
\
\tau_4 = 
\pmatrix{
{\bf 0} &  -i\sigma_3  \cr i \sigma_3 &{\bf 0}
 } ,
\\ \nonumber
&&
\tau_5 = 
\pmatrix{
I & {\bf 0} \cr {\bf 0} & -I
 } ,
\end{eqnarray}
with evident notations of $I$ and ${\bf 0}$ signifying two-by-two unit and zero
matrices respectively.
These traceless matrices form a Clifford algebra, 
$\tau_i \tau_j + \tau_j \tau_i = 2 \delta_{i,j}$.
We require the system to be $2\pi$-periodic with respect to the strength 
parameter $\lambda$, namely $e^{i 2\pi V} = 1$.
In the similar manner to $J=1/2$ case, we have,
\begin{eqnarray}
\label{e635}
V =  \frac{p}{2}+\frac{2-p}{2} \sum_{i=1}^{5} b_i \tau_i ,
\end{eqnarray}
with an integer $p$, and $\sum_i b_i^2 = 1$.
We parameterize five dimensional unit vector 
$\{ b_i \}$ as
$b_1 = \sin\gamma\cos\eta \cos\xi$, 
$b_2 = \sin\gamma\cos\eta \sin\xi$, 
$b_3 = \sin\gamma\sin\eta \cos\zeta$, 
$b_4 = \sin\gamma\sin\eta \sin\zeta$
and $ b_5 =  \cos\gamma$.
%
%
The five parameters of the system $\lambda$, $\gamma$, $\eta$, $\xi$, $\zeta$ 
form a five dimensional manifold
%
$
\{ \lambda, \gamma, \eta, \xi, \zeta \} =  S^1 \times S^4 
$.
%
The temporal evolution during one period $\Tp = 1$ is described by a unitary operator
%
$
U_\Pi = e^{-\frac{i T}{2}\tau_5} e^{-i\lambda V} e^{-\frac{i T}{2}\tau_5}
$
%
whose  eigenvalue problem
%
%
is solvable.  We have doubly-degenerate analytical solution,
\begin{eqnarray}
\label{e637}
\varepsilon_{n} 
= \frac{p}{2}\lambda + (-1)^n E_{\frac{2-p}{2}\lambda, \gamma, T} ,
\end{eqnarray}
%
%
\begin{eqnarray}
\label{e638}
&&
\left| v_{00} \right> =
\pmatrix{
e^{-i\theta_+}  \cos Q_{ \frac{2-p}{2}\lambda, \gamma, T} \cr
0 \cr
e^{-i\theta_-}  \sin \eta \sin Q_{ \frac{2-p}{2}\lambda, \gamma, T} \cr
e^{i\theta_-}  \cos \eta \sin Q_{ \frac{2-p}{2}\lambda, \gamma, T} 
}  ,
\nonumber \\
&&
\left| v_{01} \right> = 
\pmatrix{
0 \cr
e^{i\theta_+}  \cos Q_{ \frac{2-p}{2}\lambda, \gamma, T} \cr
-e^{-i\theta_-}  \cos \eta \sin Q_{ \frac{2-p}{2}\lambda, \gamma, T} \cr
e^{i\theta_-}  \sin \eta \sin Q_{ \frac{2-p}{2}\lambda, \gamma, T}  
 } ,
\nonumber \\
&&
\left| v_{10} \right> =
\pmatrix{
-e^{ -i\theta_+}  \sin \eta \sin Q_{ \frac{2-p}{2}\lambda, \gamma, T} \cr
e^{i\theta_+}  \cos \eta \sin Q_{ \frac{2-p}{2}\lambda, \gamma, T} \cr
e^{-i\theta_-} \cos Q_{ \frac{2-p}{2}\lambda, \gamma, T} \cr
0 \cr
 } ,
\\ \nonumber
&&
\left| v_{11} \right> =
\pmatrix{
-e^{-i\theta_+}  \cos \eta \sin Q_{ \frac{2-p}{2}\lambda, \gamma, T} \cr
-e^{i\theta_+} \sin \eta \sin Q_{ \frac{2-p}{2}\lambda, \gamma, T} \cr
0 \cr
e^{i\theta_-} \cos Q_{ \frac{2-p}{2}\lambda, \gamma, T} \cr
  } ,
\end{eqnarray} 
%
%
%
%
%
The structure of the eigenvalues is the same as in the previous case of $J=1/2$
apart from their double degeneracy.
With the choice $\eta=\frac{\pi}{2}$, the system is split into
two independent systems of kicked spin $J=1/2$.

The variation of eigenstates with the 
adiabatic variation of parameters is dictated by 
the Wilczek-Zee extension of
Mead-Berry connection
$A^\alpha 
= [ \left< v_{nj} \right| i \partial_\alpha \left| v_{mk} \right> ]$,
%
which transforms the state $ \left| v_{nj} \right>$ 
into $ M({C_\alpha}) \left| v_{nj} \right>$.
Here, we 
only consider the cyclic change along the $S^1$ with the variation of $\lambda$
along $C_\lambda$ : $\lambda \rightarrow \lambda+2\pi$,   
and leave the full treatment to forthcoming publication.
The matrix $M(C_\lambda) $ is given by
\begin{eqnarray}
\label{e640}
&&\!\!\!\!\!\!\!\!\!\!\!\!
M(C_\lambda) =
\cos\! \frac{(2\!-\!p)\pi}{2}
\\ \nonumber
&&\!\!\!\!
-i\sin\! \frac{(2\!-\!p)\pi}{2}
\left(\!
\sin\eta\! \left[\! \begin{array}{cc} {\bf 0} \!&\!\! -iI \\ iI \!&\!\! {\bf 0} \end{array} 
\!\right] \!+\! 
\cos\eta\!  \left[\! \begin{array}{cc} {\bf 0} \!&\! \Sigma_2 \\ \Sigma_2 \!&\! {\bf 0} \end{array} \!\right]
\!\right) .
\end{eqnarray}
For odd $p$, the first term becomes zero, and
we have the spiral holonomy displayed in 
combination of Wilczek-Zee and Berry holonomy.
%

%
%
The central finding of this work is 
that, in generating quantum holonomy,
the gauge connection has two places to act,
one of which has been neglected up to now.  
Once the assumption of single-valuedness
of states' wave function is lifted, larger class of gauge transformation 
is allowed, 
and previously mysterious spiral holonomy
can now be regarded as a legitimate member of 
the family of generalized geometric phase.
%
We may hope that the unified formulation developed in this work,
with its manifest gauge invariance, is instrumental in
further advancing our understanding of quantum holonomy.

It is not clear to us, at this point, which mathematical characterization of $A$ 
ensures $W_{m,n}$ to be permutation matrices with complex phase, 
and also which property of $A$
controls the occurrence of spiral holonomy.   
These are to be the subjects for further studies.
%

\acknowledgments
We express our gratitudes to Dr. M. Miyamoto for useful discussions and comments.  
This work has been partially supported by 
the Grant-in-Aid for Scientific Research of  Ministry of Education, 
Culture, Sports, Science and Technology, Japan
under the Grant numbers 18540384 and 18.06330.

\smallskip
This paper is dedicated to the memory of Attila Ny{\'a}.





%
%
%

\end{document}